\documentclass[lettersize,journal]{IEEEtran}
\IEEEoverridecommandlockouts

\usepackage{cite}
\usepackage{amsmath,amssymb,amsfonts}
\usepackage{breqn}
\usepackage[boxruled]{algorithm2e}
\usepackage{placeins}
\usepackage{algpseudocode}
\usepackage{graphicx}
\usepackage{textcomp}
\usepackage{xcolor}
\usepackage{float}
\usepackage{fancyhdr}
\def\BibTeX{{\rm B\kern-.05em{\sc i\kern-.025em b}\kern-.08em
    T\kern-.1667em\lower.7ex\hbox{E}\kern-.125emX}}
    
\begin{document}

\title{Critical Clearing Time Enhancement of Droop-Controlled Grid-Forming Inverters with Adaptive Function-Based Parameters\\}

\author{
\IEEEauthorblockN{
Dewan Mahnaaz Mahmud\IEEEauthorrefmark{1},
Vinu Thomas\IEEEauthorrefmark{1},
Bogdan Marinescu\IEEEauthorrefmark{2},
Micka\"{e}l Hilairet\IEEEauthorrefmark{1}
}

\IEEEauthorblockA{\IEEEauthorrefmark{1}
\textit{Nantes Université, École Centrale Nantes, LS2N, UMR 6004, F-44000 Nantes, France}
}

\IEEEauthorblockA{\IEEEauthorrefmark{2}
\textit{École Centrale Nantes, 1 Rue de la Noë, 44000 Nantes, France}
} \\ dewan-mahnaaz.mahmud@ec-nantes.fr}

\vspace{-3mm}

\maketitle
\thispagestyle{fancy}

\begin{abstract}

With the increasing penetration of renewable energy sources, grid-forming (GFM) inverters are becoming essential for voltage and frequency regulation. However, the transient stability of GFM inverter is critically affected by the current limiters that are embedded with the standard control schemes. This paper proposes a novel adaptive function to enhance the transient stability of droop-controlled GFM inverters. The proposed method autonomously adjusts the active power reference and the droop gain based on the terminal voltage of the inverter. Also, the acceleration of the phase angle is prevented, leading to the maximization of critical clearing time (CCT). The proposed method is benchmarked against two state-of-the-art GFM inverter CCT enhancement methods. Effectiveness of the proposed method is validated through electromagnetic transient (EMT) simulations in MATLAB/Simulink\textsuperscript{\textregistered}. 

\end{abstract}


\section{Introduction}

Renewable energy resources such as wind and solar are increasingly being integrated into power grids via power electronic inverters. These inverters operate as of now in grid-following mode and act as controlled current sources that are synchronized to the grid via phase-locked loops. Conventional power systems become more vulnerable to dynamic instability as the share of inverter-based resources grows and  the contribution of synchronous generators decreases. To address this, grid-forming (GFM) control has been proposed that allows the inverters to act as controlled voltage source. GFM can be represented as an internal voltage source behind an impedance that is capable of regulating grid voltage and frequency. This functionality is important in weak grids and low-inertia systems, where GFM inverters provide effective voltage and frequency support in transient conditions. The current exchanged between the inverter and the grid is determined by the impedance connecting the two voltage sources. Large voltage differences can cause currents to exceed nominal limits, necessitating the use of current-limiting mechanisms to protect the inverter \cite{b1}, \cite{b2}.

Current limiting in GFM inverters is typically accomplished with either current saturation algorithms (CSA) \cite{b3} or virtual impedance (VI)-based methods \cite{b4}. CSA saturates the current reference, whereas the VI-based method limits the current magnitude by increasing the output impedance of the inverter during transient conditions. The selection of a current limiting algorithm influences the transient stability significantly in terms of clearing time (CCT) and critical clearing angle (CCA). CCT is the maximum amount of time required to clear a fault in order for the system to remain stable. It serves as a quantitative indicator of both the severity of a large disturbance and the stability margin of the system \cite{b5} \cite{b6}. 

Authors in \cite{b7} have illustrated that VI-based limitations have better performance in transient stability, but CSA has better performance in current limitation. This analysis is done based on CCT. To improve CCT, several methods have been proposed. In \cite{b8} CCT is enhanced by optimizing the current-phase angle, $\phi_{\text{opt}}$. In \cite{b9} another method to enhance CCT is proposed by varying the droop gain. An adaptive method is proposed to increase the droop coefficient with respect to the current threshold and voltage amplitude.  In \cite{b10} the authors investigates a current limiting method that adjusts the active power set point when inverter current surpasses a chosen threshold value. Power setpoint is then reduced according to the proportional gain. In \cite{b11} a method to enhance transient angle stability is proposed based on a power compensation loop. An additional torque is generated in the power synchronization loop (PSL) based on a proportional gain that amplifies the error between the grid voltage and rated voltage. All these methods can indeed enhance CCT. However, these methodologies require parameters to tune or to solve an optimization problem.  Also, improper tuning of the proportional controller can degrade the stability of the system.  

\begin{figure*}[t!]
\centering
\includegraphics[width=\linewidth]{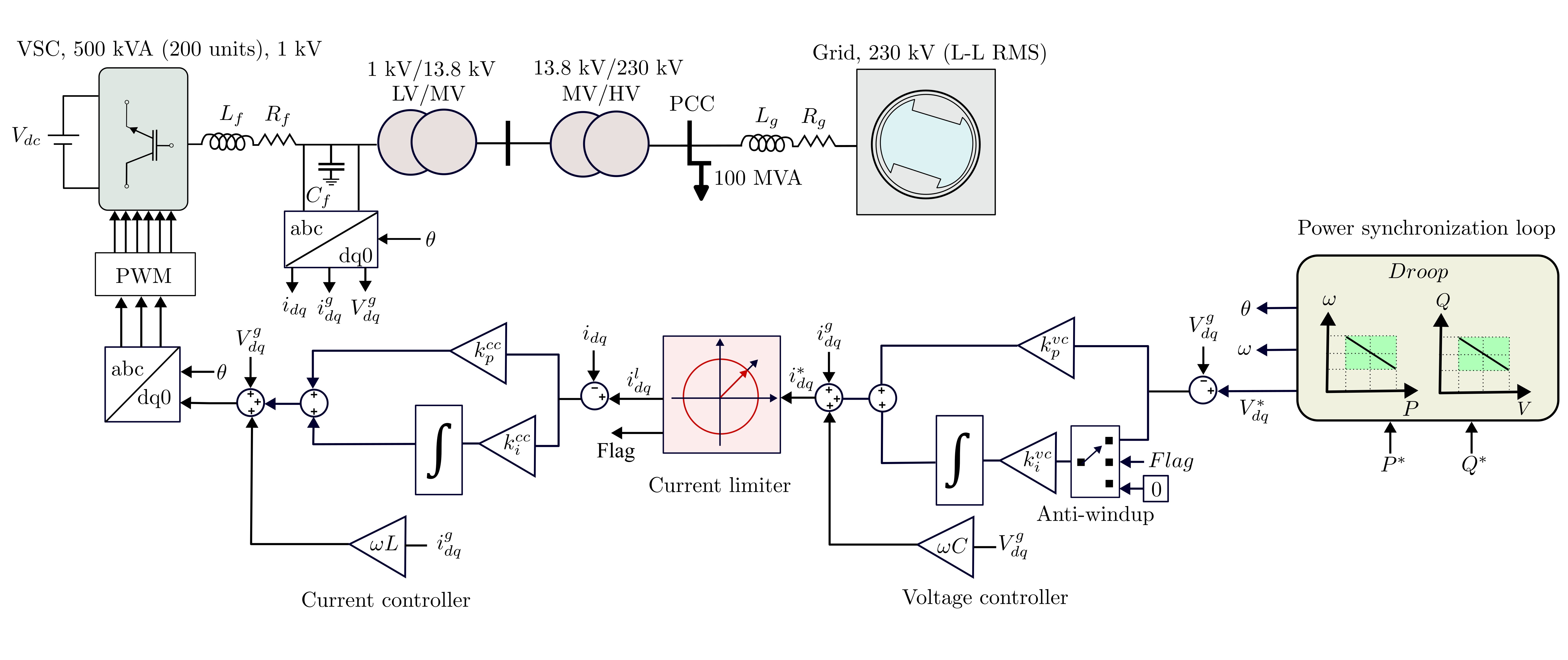}
\caption{System architecture for standard droop-based GFM control}
\label{fig:Sys}
\end{figure*}

To address these limitations, a novel adaptive function is proposed. This function modifies the synchronization loop of the droop-controlled GFM inverters to enhance CCT. The proposed adaptive function adjusts the power reference and droop gain with respect to the terminal voltage of the inverter. As a result, the inverter is prevented from the acceleration of the synchronization angle during transient events in order to maximize CCT. Finally, the proposed method operates autonomously using local measurements without requiring any additional parameters to tune or external supervisory signals.

This paper is structured as follows: Section \ref{section:2} describes the system description. Critical clearing time estimation and post-fault resynchronization are derived in \ref{section:3}. Section \ref{section:4} presents the proposed control, followed by the case studies in section \ref{section:5}. Finally, in section \ref{section:6}, conclusion and future work are summarized.  

\section{System description} \label{section:2}

The system architecture with the control diagram is depicted in Fig.\ref{fig:Sys}. The inverter is equipped with an LC filter in which the inductance of the filter is $L_f$, the capacitance is $C_f$, and the resistance is $R_f$. The grid is represented as an ideal voltage source $V_g$ in series with a line impedance of $R_g$ and $L_g$. The GFM inverter is controlled by an outer voltage control loop and an inner current control loop. Proportional-integral (PI) controllers are used in both of the loops. The synchronization angle $\theta$ is generated from the droop-based PSL. A circular current limiter is used to protect the GFM inverter from overcurrent. 

In this system in Fig.\ref{fig:Sys}, each GFM represents an aggregated model of 200 commercial inverter modules. Each module is rated as 500 kVA which is connected with an LV-MV and MV-HV transformer. More details of the aggregated inverter model can be found in \cite{b10}. A load of 100 MVA is connected at the point of common coupling (PCC) with the voltage level of 230 kV. The parameters of the system are shown in Table \ref{table:system_params}.

\section{Post-fault resynchronization and critical clearing time estimation} \label{section:3}

When the circular current limiter is activated, the GFM inverter operates as a controlled current source. The magnitude of the injected current is constrained to $I_{\max}$. The expression of active power is shown in \eqref{eq:P_lim}.

\begin{equation}
P_f = V_g I_{\max} \cos(\delta - \varphi)
\label{eq:P_lim}
\end{equation}

In eq. \ref{eq:P_lim}, $P_f$ denotes the instantaneous active power delivered to the grid during current-limited operation, and $V_g$ is the grid voltage magnitude. The angle $\delta$ represents the phase angle difference, while $\varphi$ is the phase angle of the current. 

During transients, the mismatch between the reference power $P^*$ and the delivered power $P_f$ governs the angle dynamics shown in eq. \ref{eq:swing_fault}:

\begin{equation}
\frac{d\delta}{dt} = m_p \omega_b \left(P^* - P_f\right)
\label{eq:swing_fault}
\end{equation}

Critical clearing angle $\delta_{\mathrm{cca}}$ is determined using the equal area criterion on the saturated  $P(\delta)$ curve. The corresponding critical clearing time (CCT) is obtained by integrating the angle dynamics from the pre-fault angle $\delta_0$ to $\delta_{\mathrm{cca}}$ as given in eq. \ref{eq:CCTF}:

\begin{equation}
t_{\mathrm{cc}} 
= \int_{\delta_0}^{\delta_{\mathrm{cca}}}
\frac{d\delta}{m_p \omega_b \left(P^* - P_f(\delta)\right)} 
\label{eq:CCTF}
\end{equation}

The detailed analytical derivation of the CCT is provided in \cite{b12}.

\begin{table}[t]
\centering
\caption{System Parameters}
\label{table:system_params}
\begin{tabular}{|l|l|}
\hline
\textbf{System base} &  \\
\hline
Base power, \( S_b \) & 100 MVA \\
Base voltage, \( v_b \) & 230 kV \\
Base frequency, \( f_b \) & 50 Hz \\
\hline

\textbf{MV/HV transformer (13.8/230 kV)} &  \\
\hline
Rated power, $S_T$ & 210 MVA \\
Winding resistance, $(R_1, R_2)$ & 0.0027 pu \\
Winding inductance, $(L_1, L_2)$ & 0.08 pu \\
Magnetizing branch, $(R_m, L_m)$ & 500, 500 pu \\
\hline

\textbf{LV/MV transformer (1/13.8 kV)} &  \\
\hline
Rated power, $S_T$ & 1.6 MVA \\
Winding resistance, $(R_1, R_2)$ & 0.0073 pu \\
Winding inductance, $(L_1, L_2)$ & 0.018 pu \\
Magnetizing branch, $(R_m, L_m)$ & 347, 156 pu \\
\hline

\textbf{Single-unit inverter} &  \\
\hline
Rated power, $S_{\mathrm{vsc}}$ & 500 kVA \\
Filter $(R, L, C)$ & $0.001~\Omega,\,200~\mu\text{H},\,300~\mu\text{F}$ \\
Number of inverter modules, $n$ & 200 \\
DC voltage, \( v_{dc} \) & 2.44 kV \\
AC voltage, \( v_{rms} \) & 1 kV \\
Droop coefficient, \( m_p \) & 0.05 \\
\hline
\end{tabular}
\vspace{-2mm}
\end{table}

\section{Proposed adaptive function} \label{section:4}

\begin{figure}
    \centering
    \includegraphics[width=0.7\linewidth]{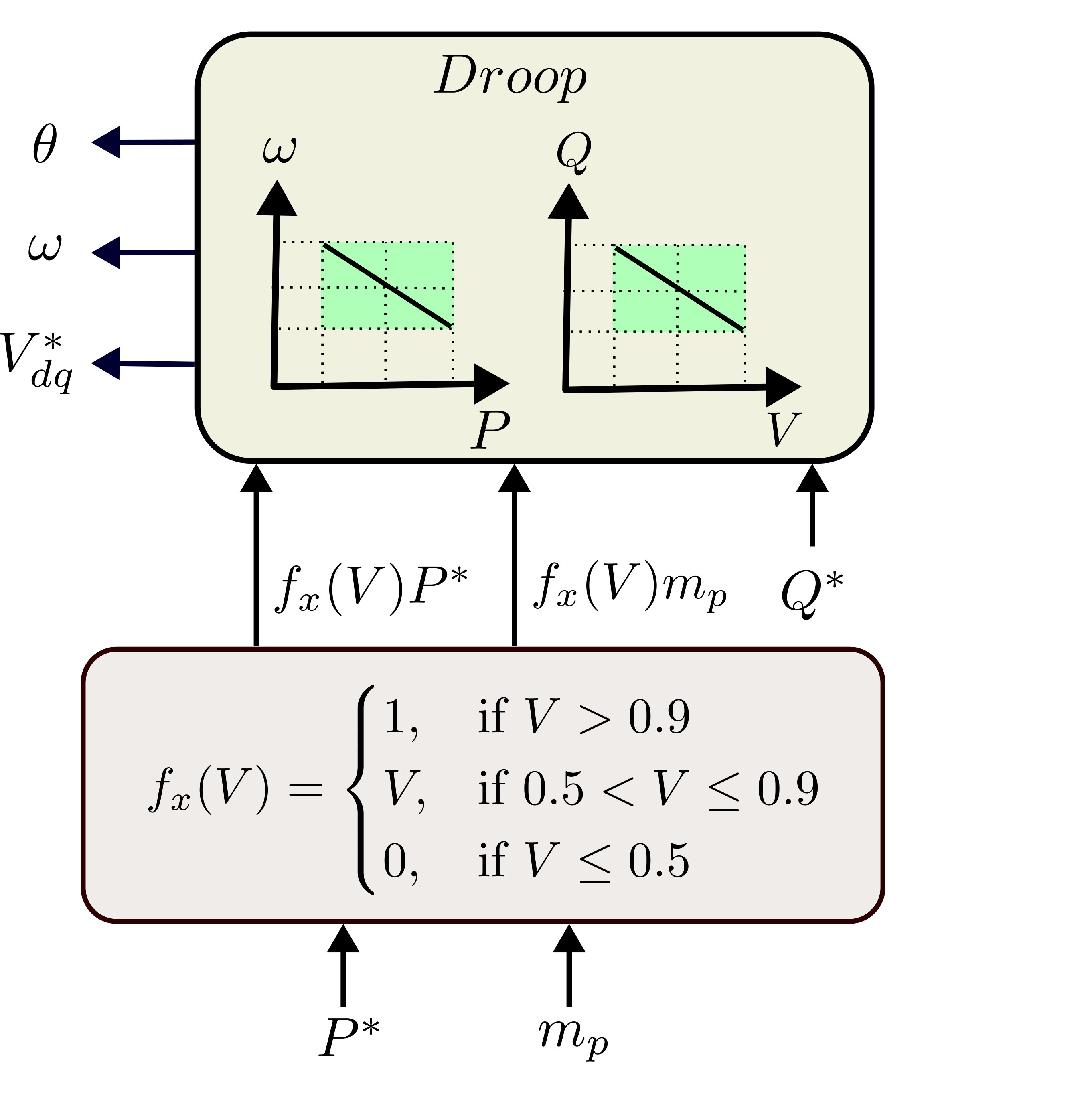}
    \caption{Adaptive function-based droop control scheme}
    \label{fig:PMDC}
\end{figure}

\begin{figure}
    \centering
    \includegraphics[width=0.9\linewidth]{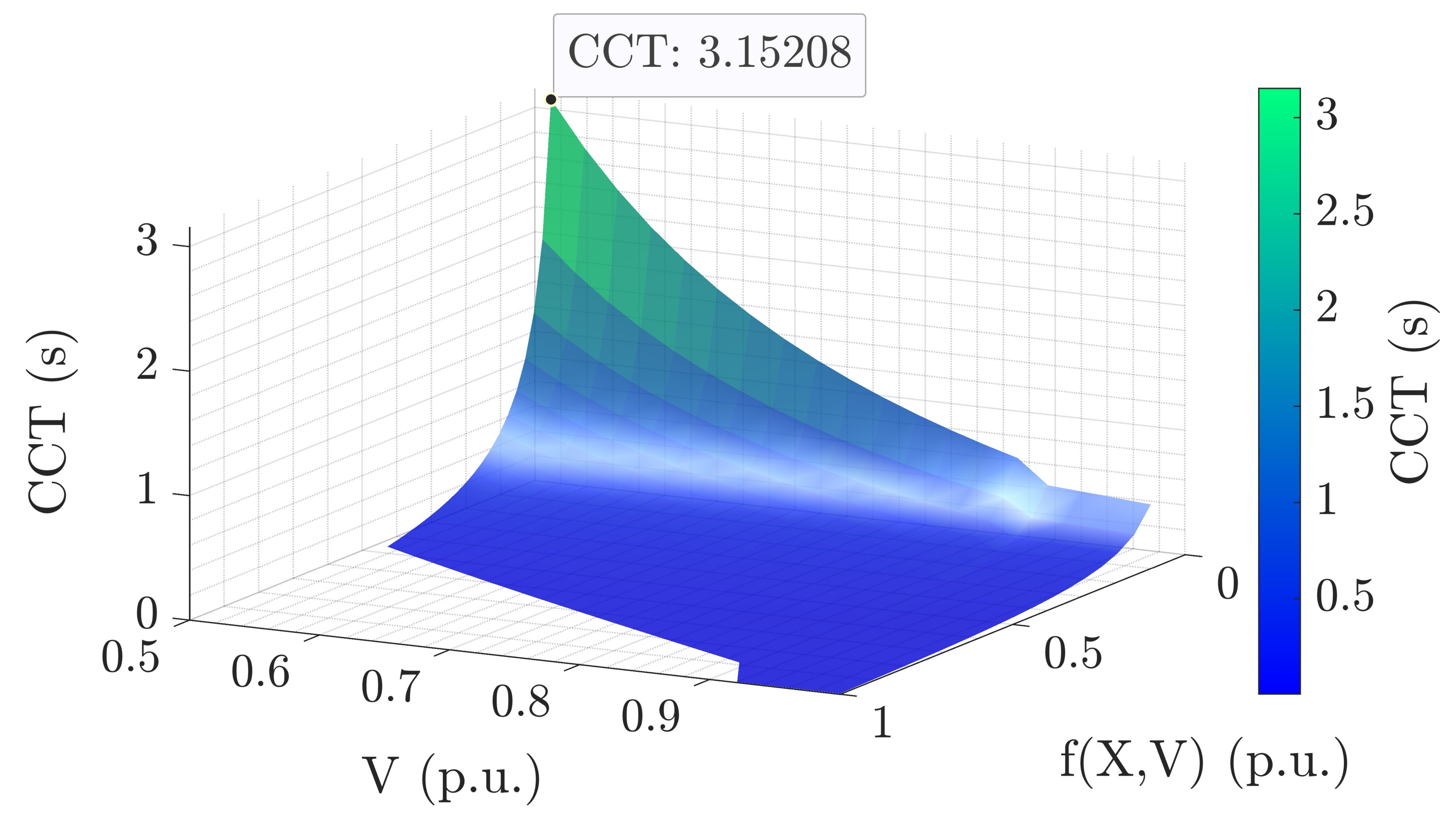}
    \caption{Surface plot of the proposed function}
    \label{fig:CCTSF}
\end{figure}

The basic idea of the proposed adaptive function is to adjust the parameters of the droop-based power synchronization loop based on the terminal voltage of the inverter. This approach reduces angle acceleration during large disturbances and directly enhances the critical clearing time (CCT). The adaptive function is shown in \eqref{eq:PM}. The proposed function \( f_x(V) \) modifies \( x \) according to the instantaneous value of \( V \), where \( V \) determines whether the output of the function should remain unchanged or is to be adjusted. Here, \( x \) represents parameters to be adapted: droop coefficien, \( m_p \) and power reference, \( P^* \) and  \( V \) represents the instantaneous terminal voltage of the inverter. During voltage dips, \( f_x(V) \) reduces the value of \( x \), limiting angle acceleration and improving system CCT.

\begin{equation}
    f_x(V) =
    \begin{cases} 
    1, & \text{if } V > 0.9 \\
    V, & \text{if } 0.5 < V \leq 0.9 \\
    0, & \text{if } V \leq 0.5
    \end{cases}
\label{eq:PM}
\end{equation}

Using this proposed function, \( P^* \) and \( m_p \) are adaptively updated. The modified droop equation is given in \eqref{eq:NDE}. According to the CCT expression in \eqref{eq:CCTF}, CCT increases as the term \( m_p \omega_b (P^* - P_f) \) decreases. Reducing \( m_p \) and \( P^* \) during voltage dips limits angle acceleration and slows the divergence of the internal angle of the inverter. 

\begin{equation}
\omega = \omega^*
+ [\underbrace{m_p f_x(V)}_{\substack{\text{adaptive droop} \\ \text{coefficient}}}]
[\underbrace{P^* f_x(V)}_{\substack{\text{adaptive power} \\ \text{reference}}}
- P]
\label{eq:NDE}
\end{equation}

For the terminal voltage higher than 0.9 pu, the adaptive function equals unity, $f_x(V)=1$. So, both \( P^* \) and \( m_p \) remain at their nominal values. Under these conditions, the inverter operates in its steady state with no adaptive modification is introduced. When the voltage at the inverter terminal is between 0.5 and 0.9 pu, the adaptive function \( f_x(V) = V \), that proportionally reduces both \( P^* \) and \( m_p \). This reduction effectively lowers the active power injection and modifies the slope of the droop. This adaptive adjustment is preventing excessive acceleration of the internal angle, which commonly occurs when the inverter continues to inject predefined active power under voltage dips. 

For severe and nearby faults, the terminal voltage drops might drop below 0.5 pu. In this case, the adaptive function output becomes zero. Thus, active power injection and corresponding accelerating torque are effectively nullified. CCT approaches theoretically to infinity. In reality, it means the inverter remains synchronized with the system till the fault is cleared. The implementation of the adaptive function is shown in Fig. \ref{fig:PMDC}. Without the proposed adaptive function, a larger power mismatch would increase acceleration and push the inverter closer to the maximum critical clearing angle (CCA). By adjusting \( P^* \) and \( m_p \), angle acceleration is prevented and CCT is enhanced. 

In the previous methods that address CCT enhancement, droop gain variation in \cite{b9} improves CCT by decreasing the droop gain based on the current or voltage thresholds. However, it remains sensitive to small deviations that can be unwanted. In \cite{b10}, active power setpoint is reduced by current threshold and a proportional gain to increase CCT. This method requires an extra gain to be tuned. Improper tuning can degrade system performance, even leading to instability. In \cite{b8}, CCT is enhanced by optimizing the current-phase angle, ($\phi_{\text{opt}}$). This method requires solving an optimization problem and no longer complies with the latest technical requirements for GFM inverters \cite{b13}. The proposed adaptive function in \eqref{eq:PM} addresses this limitation and maximizes CCT. This eliminates the need for an extra gain to tune or sensitivity to small deviations and complies with the technical requirement.

CCT enhancement is shown in the surface plot of fig. \ref{fig:CCTSF}. From the surface plot, it is observed that CCT is significantly increased to 3.15 s. Note that, CCT of 3.15 s is achieved specifically in the voltage range \(0.5 < V \leq 0.9\). When the terminal voltage drops further, the output of the adaptive function \(f_x(V) = 0\). Therefore, the surface plot is only meaningful for \(V > 0.5\). This value provides a stability margin that is considerably higher than the minimum specified by the current grid code \cite{b14} as of now.

\section{Case studies} \label{section:5}

\subsection{CCT validation}

\begin{figure}
    \centering
    \includegraphics[width=1.0\linewidth]{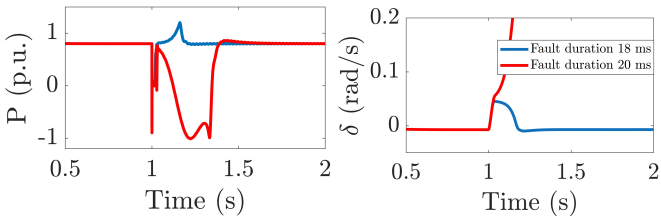}
    \caption{Validation of CCT}
    \label{fig:CCTV}
\end{figure}

Theoretical estimation of CCT is validated using electromagnetic transient (EMT) simulation, which is done in MATLAB/Simulink\textsuperscript{\textregistered}. Fig. \ref{fig:CCTV} presents the system response to symmetrical three-phase bolted faults applied at the PCC for two distinct durations: 18 ms and 20 ms. The phase angle difference shows the fundamental difference between the two cases. 

When the duration of the fault is 20 ms, unstable behavior is observed after the fault is cleared. Phase angle $\delta(t)$ continues to increase rather than returning to equilibrium. As a result, the inverter starts absorbing active power. This behavior indicates loss of synchronism. The inverter starts to operate at a new equilibrium point once the synchronism is lost, which is unwanted. In contrast, when an 18 ms fault is applied, $\delta(t)$ exhibits an oscillation with the deviation reaching approximately 0.05 rad/s before returning to a steady-state equilibrium. This behavior indicates that the stability is maintained as $\delta(t)$ returns to the pre-fault operating condition. During the fault period, the active power collapsed, and there is a mismatch between the $P$ and $P^*$. This is the reason to increase the phase difference between the inverter and the grid. When the fault is cleared, $P$ is recovered with an overshoot that provides the necessary deceleration that enables the system to recover. 

This behavior confirms that 18 ms represents the CCT obtained from the simulation. Using the derived formula in \eqref{eq:CCTF} CCT is obtained at 19.4 ms, which is close to the EMT simulation. The difference is due to the simplifying assumptions that neglect the dynamics of the LV/MV and MV/HV transformers, also the current and voltage control loop is considered to have a unity gain. Thus, CCT can be expressed by eq. \ref{eq:CCTF}.

\subsection{$40\%$ voltage sag at PCC}

\begin{figure}
    \centering
    \includegraphics[width=0.9\linewidth]{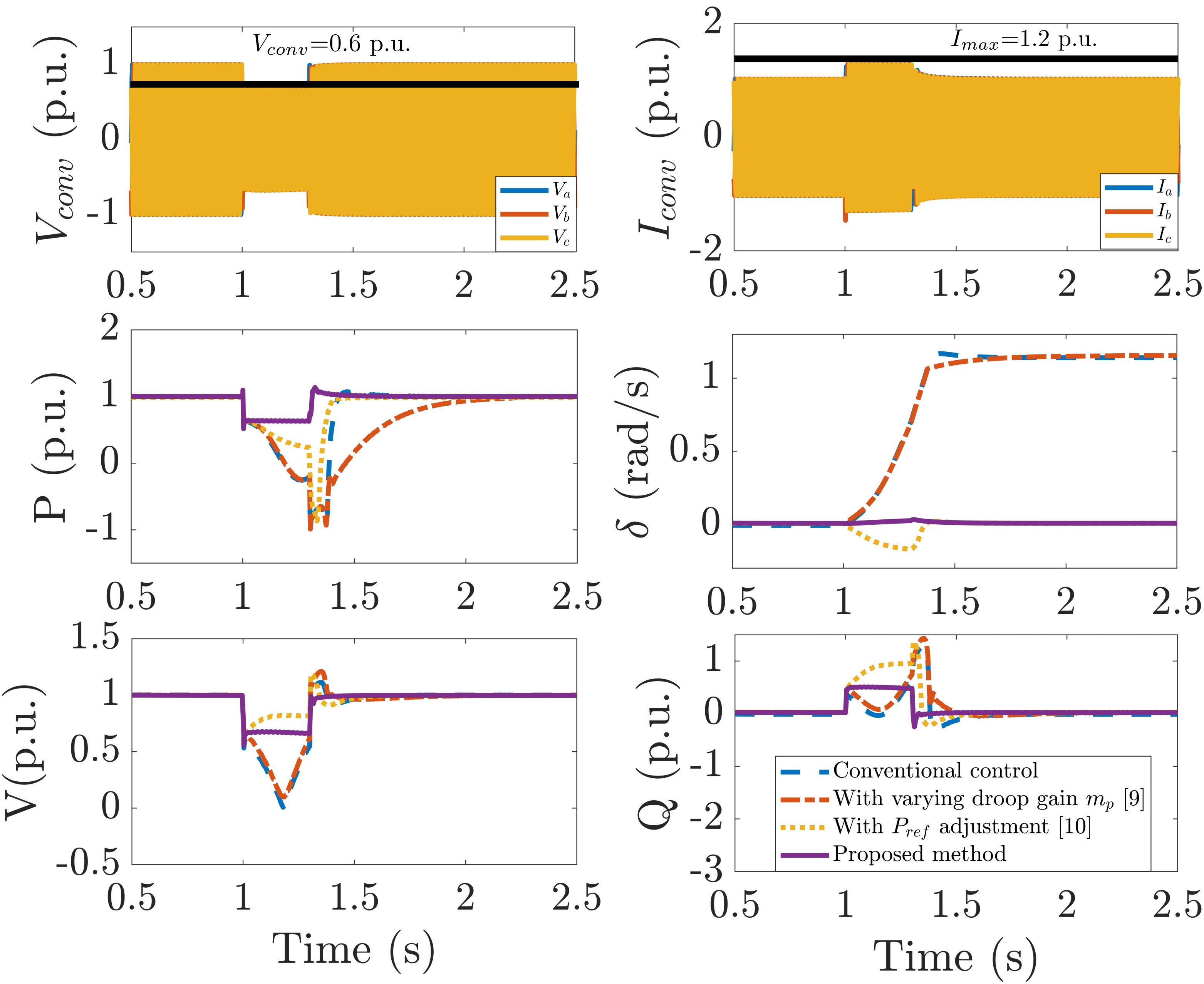}
    \caption{40\% voltage sag at PCC for 400 ms.}
    \label{fig:GVC_40_PCC}
\end{figure}

To validate the proposed method, \(40\%\) voltage sag (represents a faulted case of non-zero resistance) at the PCC is applied for a duration of 400 ms. Fig. \ref{fig:GVC_40_PCC} illustrates the inverter response under this condition. Since the inverter terminal voltage lies in the range \(0.5 < V \leq 0.9\), the power transfer between the grid and the inverter is reduced. However, as the \(P^*\) remains unchanged, the phase angle difference between the converter and the grid continues to increase. As a result, synchronism is lost when the grid voltage recovers. 

In addition, the PCC voltage tends to collapse because sufficient reactive power is not injected due to the absence of active power curtailment. This unstable behavior is observed with the conventional control as well as the method reported in \cite{b9}. In contrast, the approach in \cite{b10} exhibits different dynamics, as  \(P^*\) is curtailed based on a current threshold using a proportional gain. As the proportional gain is poorly tuned, the active power is reduced too aggressively causing the inverter to absorb active power as the phase angle becomes negative which is an abnormal and undesirable operating condition.

Proposed method avoids this issue, as \(P^*\) is curtailed smoothly based on the measured voltage in the range \(0.5 < V \leq 0.9\). This prevents PCC voltage collapse and preserves the phase angle difference by keeping the delivered power aligned with the reference, thereby maintaining stable system operation during and after the voltage sag.

\subsection{Bolted fault}

\begin{figure}
    \centering
    \includegraphics[width=0.9\linewidth]{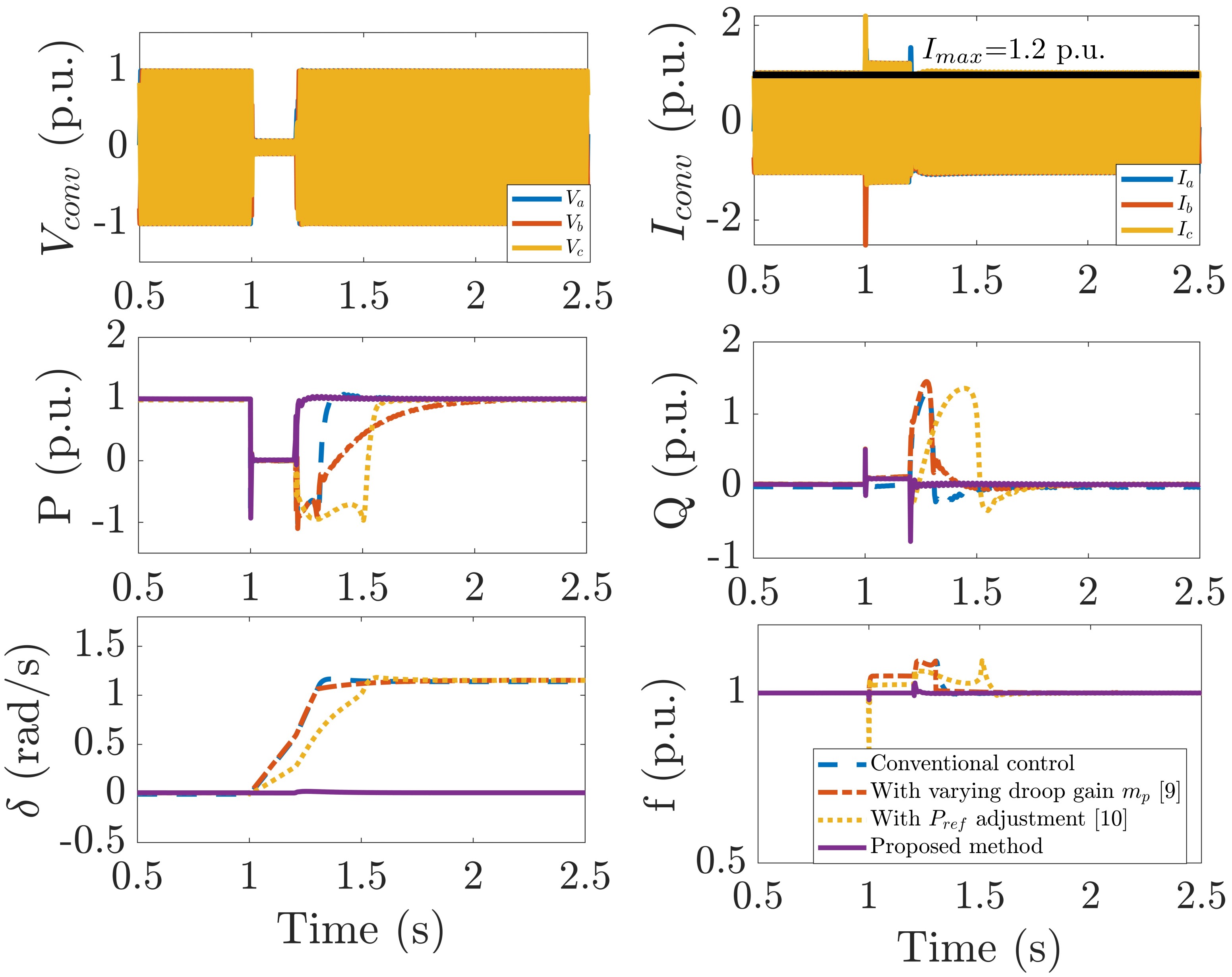}
    \caption{Bolted fault at PCC for 200 ms}
    \label{fig:BF_PCC}
\end{figure}

Further to validate the proposed method, a three-phase bolted fault with zero resistance is applied at the PCC of the system for 200 ms. Fig. \ref{fig:BF_PCC} shows the response of the proposed method. When the fault happens, voltage at PCC drops to zero. As a consequence, there is no power transfer between the inverter and grid. It results in the phase angle difference between the converter and the grid to increase as the setpoint of the inverter is still high. When the fault is cleared, the method proposed in \cite{b9} and \cite{b10} fails to synchronize, whereas the proposed method re-synchronizes as soon as the fault is cleared. This happens because during faulted conditions the inverter is operating at nominal $P^*$. It is unwanted because the angle of the power synchronization loop keeps increasing, which pushes the phase angle of the inverter near the $\delta_{CCA}$. Once it surpasses $\delta_{CCA}$ , the inverter loses synchronism and resynchronizes at a new equilibrium point. But in the proposed method, the $P^*$ and $m_{p}$ is reduced during the bolted fault. This results in minimum deviation of the phase angle difference. Hence, the synchronization stability is preserved when the bolted fault is cleared. 

\begin{figure}
    \centering
    \includegraphics[width=0.85\linewidth]{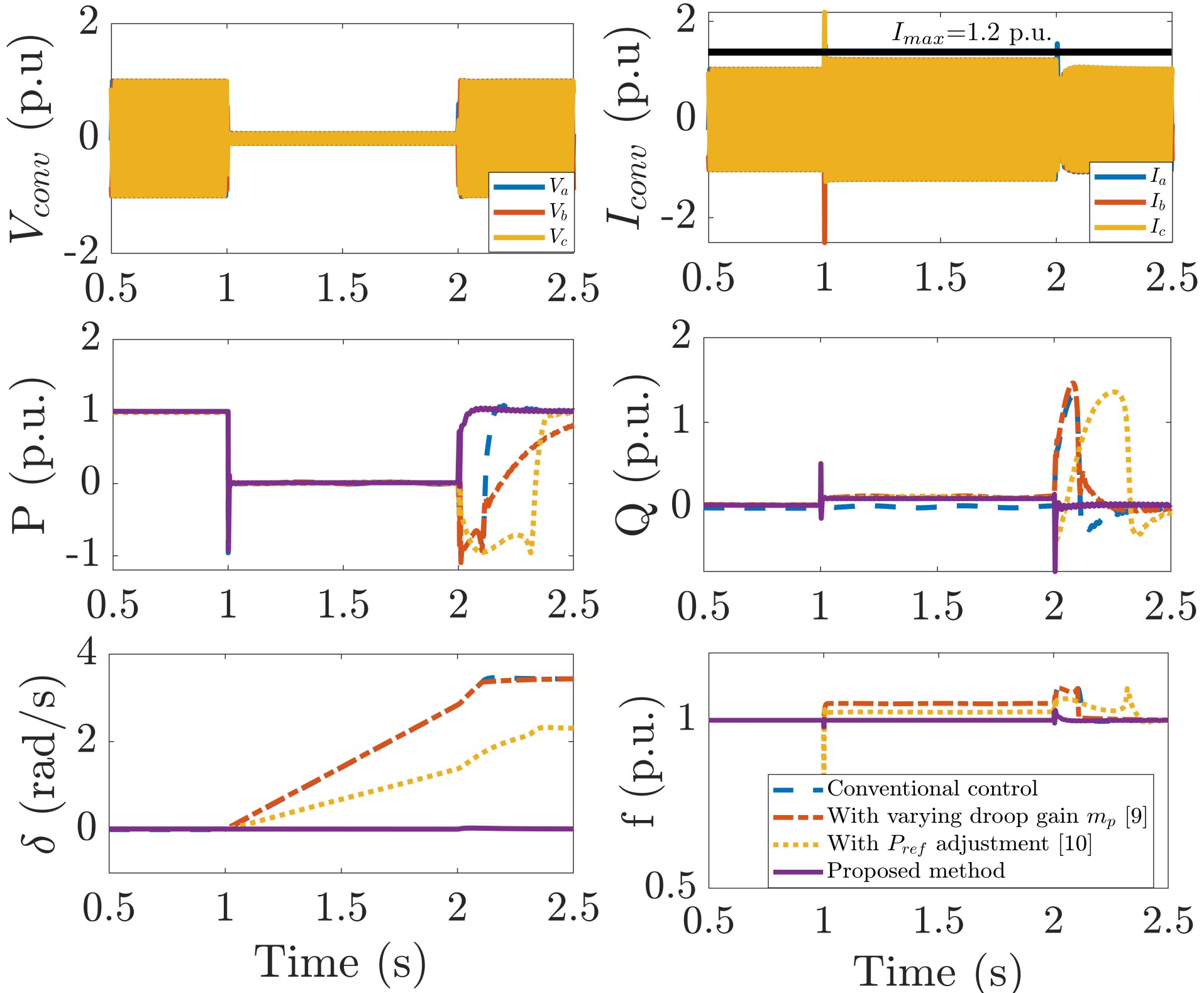}
    \caption{Bolted fault at PCC for 1s}
    \label{fig:BF_PCC_1s}
\end{figure}

To further test the effectiveness of the proposed control, a bolted fault of 1s (represents an extreme case) is applied at PCC. Fig. \ref{fig:BF_PCC_1s} shows the response for this extreme case. The proposed method is able to keep synchronized with the grid as $P^*$ and $m_{p}$ is decreased according to the function. Whereas the other methods fail to resynchronize as they surpass the $\delta_{CCA}$. In the proposed method, the phase angle does not surpass  $\delta_{CCA}$ as a result, the system remains stable. So, it can be concluded that the proposed method enhances CCT.

\section{Conclusion} \label{section:6}

This paper introduces a novel adaptive function-based power synchronization loop that enhances the transient stability of droop-controlled GFM inverters. This adaptive function modifies the synchronization loop to automatically adjust the power reference and droop gain with respect to the instantaneous terminal voltage of the inverter. This method effectively prevents phase angle acceleration during faults, extending the CCT without requiring parameter tuning or external supervisory control.  Theoretical analysis and EMT simulation demonstrates that the proposed method is effective in terms of CCT maximization. Future work will be on the experimental validation and extension in the multi-inverter systems.


\begin{thebibliography}{00}
\bibitem{b1} N. Baeckeland, D. Chatterjee, M. Lu, B. Johnson, and G.-S. Seo, ``Overcurrent limiting in grid-forming inverters: A comprehensive review and discussion,'' \textit{IEEE Trans. Power Electron.}, 2024.

\bibitem{b2} S. H. Khan, M. Z. Lazkano, P. Izurza, A. Sanchez-Ruiz, J. C. Acena, and J. Arza, ``Synchronization stability of a grid forming converter under the effect of current limit in voltage dips with VI based current limiting method: Analysis and solution,'' in \textit{Proc. 24th Eur. Conf. Power Electron. Appl. (EPE'22 ECCE Europe)}, 2022, p. P--1.

\bibitem{b3} N. Bottrell and T. C. Green, ``Comparison of current-limiting strategies during fault ride-through of inverters to prevent latch-up and wind-up,'' \textit{IEEE Trans. Power Electron.}, vol. 29, no. 7, pp. 3786--3797, 2013.

\bibitem{b4} A. D. Paquette and D. M. Divan, ``Virtual impedance current limiting for inverters in microgrids with synchronous generators,'' \textit{IEEE Trans. Ind. Appl.}, vol. 51, no. 2, pp. 1630--1638, 2014.

\bibitem{b5} X. Lyu, W. Du, S. M. Mohiuddin, S. P. Nandanoori, and M. Elizondo, ``Criteria for grid-forming inverters transitioning between current limiting mode and normal operation,'' \textit{IEEE Trans. Power Syst.}, vol. 39, no. 4, pp. 6107--6110, 2024.

\bibitem{b6} Y. Zhang, F. Liu, and Q. Guo, ``Critical clearing time sensitivity of power systems with high power electronic penetration,'' \textit{iEnergy}, 2025.

\bibitem{b7} T. Qoria, F. Gruson, F. Colas, X. Kestelyn, and X. Guillaud, ``Current limiting algorithms and transient stability analysis of grid-forming VSCs,'' \textit{Electr. Power Syst. Res.}, vol. 189, p. 106726, 2020.

\bibitem{b8} E. Rokrok, T. Qoria, A. Bruyere, B. Francois, and X. Guillaud, ``Transient stability assessment and enhancement of grid-forming converters embedding current reference saturation as current limiting strategy,'' \textit{IEEE Trans. Power Syst.}, vol. 37, no. 2, pp. 1519--1531, 2021.

\bibitem{b9} T. Qoria, F. Gruson, F. Colas, G. Denis, T. Prevost, and X. Guillaud, ``Critical clearing time determination and enhancement of grid-forming converters embedding virtual impedance as current limitation algorithm,'' \textit{IEEE J. Emerg. Sel. Topics Power Electron.}, vol. 8, no. 2, pp. 1050--1061, 2019.

\bibitem{b10} A. Tayyebi, D. Gross, A. Anta, F. Kupzog, and F. Dorfler, ``Frequency stability of synchronous machines and grid-forming power converters,'' \textit{IEEE J. Emerg. Sel. Topics Power Electron.}, vol. 8, no. 2, pp. 1004--1018, 2020.

\bibitem{b11} Z. Shuai, C. Shen, X. Liu, Z. Li, and Z. J. Shen, ``Transient angle stability of virtual synchronous generators using Lyapunov’s direct method,'' \textit{IEEE Trans. Smart Grid}, vol. 10, no. 4, pp. 4648--4661, 2018.

\bibitem{b12} X. Lyu, W. Du, S. M. Mohiuddin, and Y. Cheng, ``Critical clearing time for droop-controlled grid-forming inverters with circular current limiting and virtual impedance current limiting,'' \textit{IEEE Trans. Power Syst.}, 2025.

\bibitem{b13} European Network of Transmission System Operators for Electricity (ENTSO-E), ``Grid Forming Capability of Power Park Modules: Report on Technical Requirements,'' Tech. Rep., Oct. 3, 2025.

\bibitem{b14} RG-CE System Protection \& Dynamics Sub Group, ``Determining Generator Fault Clearing Time for the Synchronous Zone of Continental Europe - Version 1.0,'' Tech. Rep., Feb. 3, 2017.
\end{thebibliography}
\end{document}